# THE SHADOW OF LIGHT: LORENTZ INVARIANCE AND COMPLEMENTARITY PRINCIPLE IN ANOMALOUS PHOTON BEHAVIOUR

**F. Cardone[a], R. Mignani[b,c,*] , W. Perconti[d], A. Petrucci[b], R. Scrimaglio[d]**

[a] Istituto per lo Studio dei Materiali Nanostrutturati (ISMN-CNR), via dei Taurini 19, 00185 Roma, Italy

[b] Dipartimento di Fisica "E. Amaldi", Università degli Studi "Roma Tre", via della Vasca Navale. 84, 00146 Roma, Italy

[c]INFN, Sezione di Roma III

[d]Dipartimento di Fisica, Università degli Studi de L'Aquila, via Vetoio 1, 67010 Coppito, L'Aquila , Italy

***Abstract***

We review the results of two double-slit-like experiments in the infrared range, which evidence an anomalous behaviour of photon systems under particular (energy and space) constraints. These outcomes (independently confirmed by crossing photon beam experiments in both the optical and the microwave range) apparently rule out the Copenhagen interpretation of the quantum wave, i.e. the probability wave, by admitting an interpretation in terms of the Einstein-de Broglie-Bohm hollow wave for photons. Moreover, these experiments support the interpretation of the hollow wave as a deformation of the Minkowski space-time geometry. We stress the implications of these experimental results and of their interpretation for the concept of action-at-a-distance, Einstein's relativistic correlation and Bohr's principle of complementarity.

## 1. Introduction

The purpose of this paper is discussing the experimental foundations and the theoretical implications of an hypothesis we put forward recently [1,2], namely that the wave associated to a quantum object (to be meant according to the Einstein-de Broglie-Bohm interpretation, i.e. as a hollow – or pilot – wave) is a deformation of the Minkowskian space-time geometry. This hypothesis establishes a possible connection between the two seemingly unrelated questions of the real nature of the quantum wave in Quantum Mechanics, and of the possible breakdown of local Lorentz invariance in Special Relativity.

Wave-particle duality (in particular, the interpretation of the wave nature of quantum entities) is a long-debated issue in Quantum Mechanics, but far from being resolved. The wave associated to a quantum object is commonly regarded as a probability wave, according to what is usually referred to as the Copenhagen interpretation[1], and hence it conveys no physical properties.

This interpretation is thoroughly antithetic to that advocated by Einstein, De Broglie and Bohm, which regards the quantum wave as real, intimately bound to the quantum entity and moving along with it, but unable to carry energy and momentum (hollow or ghost wave). At present, the theoretical hypothesis of a real, hollow wave is necessary in order to correctly interpret some experiments, which evidence the wave-corpuscle duality

[*] Corresponding author.
*E-mail address:* mignani@fis.uniroma3.it (R. Mignani).

[1] Actually, the probabilistic interpretation of the wave function was due to M. Born of the Göttingen school.

(although they do not provide any direct evidence of hollow waves). Nevertheless, it is possible to attain indirect proof of hollow waves thanks to the influence they have on the behaviour and occurrences of events that overlap in space-time, like interference phenomena.

In the framework of Special Relativity, it is a controversial topic (from both the theoretical and the experimental side) as to whether the validity of local Lorentz invariance (LLI) is preserved at any length or energy scale. The experiments aimed at testing LLI can be roughly classified in three groups:

i) Michelson-Morley type experiments, which test isotropy of the round-trip speed of light;
ii) experiments which test isotropy of the one-way speed of light, based on atomic spectroscopy and atomic timekeeping;
iii) Hughes-Drever type experiments, testing isotropy of nuclear energy levels.

All such experiments set upper bounds on the degree of violation of LLI.

A possible threshold value for a breakdown of local Lorentz invariance in electromagnetic interactions was obtained by two of the present authors (F.C. and R.M) [3] by the formalism of Deformed Special Relativity (DSR). Such a generalization of Special Relativity is based on a “deformation” of the Minkowski space, namely a space-time endowed with a metric whose coefficients depend on the energy of the investigated processes.

More precisely, the analysis of the Cologne experiment [4] (superluminal sub-cutoff propagation in waveguide), and of the Florence one [5] (superluminal propagation in air), carried out by means of the DSR formalism, brought about upper threshold values both in energy and in space for the electromagnetic breakdown of LLI. These values are $E_0 = 4.5\ \mu V$ and $l_0 = 9\ cm$ [3,6] respectively.

The connection between the quantum wave (according to the Einstein-De Broglie-Bohm interpretation) and the breakdown of local Lorentz invariance (described by the DSR formalism) lies in the hypothesis (we put forward in our papers [1,2]) that *a hollow wave is nothing but a local deformation of space-time geometry.* By a metaphoric image we may picture the local deformed space-time, which is intimately bound to each photon, as *the shadow of the photon*. It is immaterial, like a shadow (since it carries neither energy nor momentum) and it can reach space regions far from the photon, exactly as a shadow fills space regions far from the body that casts it. It has been our target to study and pinpoint the possible effects of such a shadow of light in different experimental conditions.

Thus, in order to detect the effects of hollow waves and hence test our hypothesis, we carried out two optical experiments [1,2], of the double-slit type, designed according to the energy and space threshold behaviour of the LLI breakdown. The experimental set-up, measurement procedure and results of these two experiments are illustrated in Subsect.2.1. Subsect.2.2 briefly reports the results of two interference experiments involving crossed photon beams, in the microwave and in the optical range, which confirm our findings. Sect. 3 is devoted to the analysis of the results obtained in the two experiments, their interpretation in terms of the hollow waves conceived as space-time deformations, and their implications, including failure of relativistic correlation and invalidation of Bohr’s complementarity.

## 2. Experimental evidence for anomalous photon behaviour

### *2.1. Double-slit like experiments*

#### *2.1.1. First experiment*

The experiments we carried out were optical ones, in the infrared range, of the double-slit type. Let us briefly report the main features and results of the first experiment, carried out at L'Aquila in 2002 [1].

The employed apparatus (schematically depicted in Fig.1) consisted of a Plexiglas box with wooden base and lid.

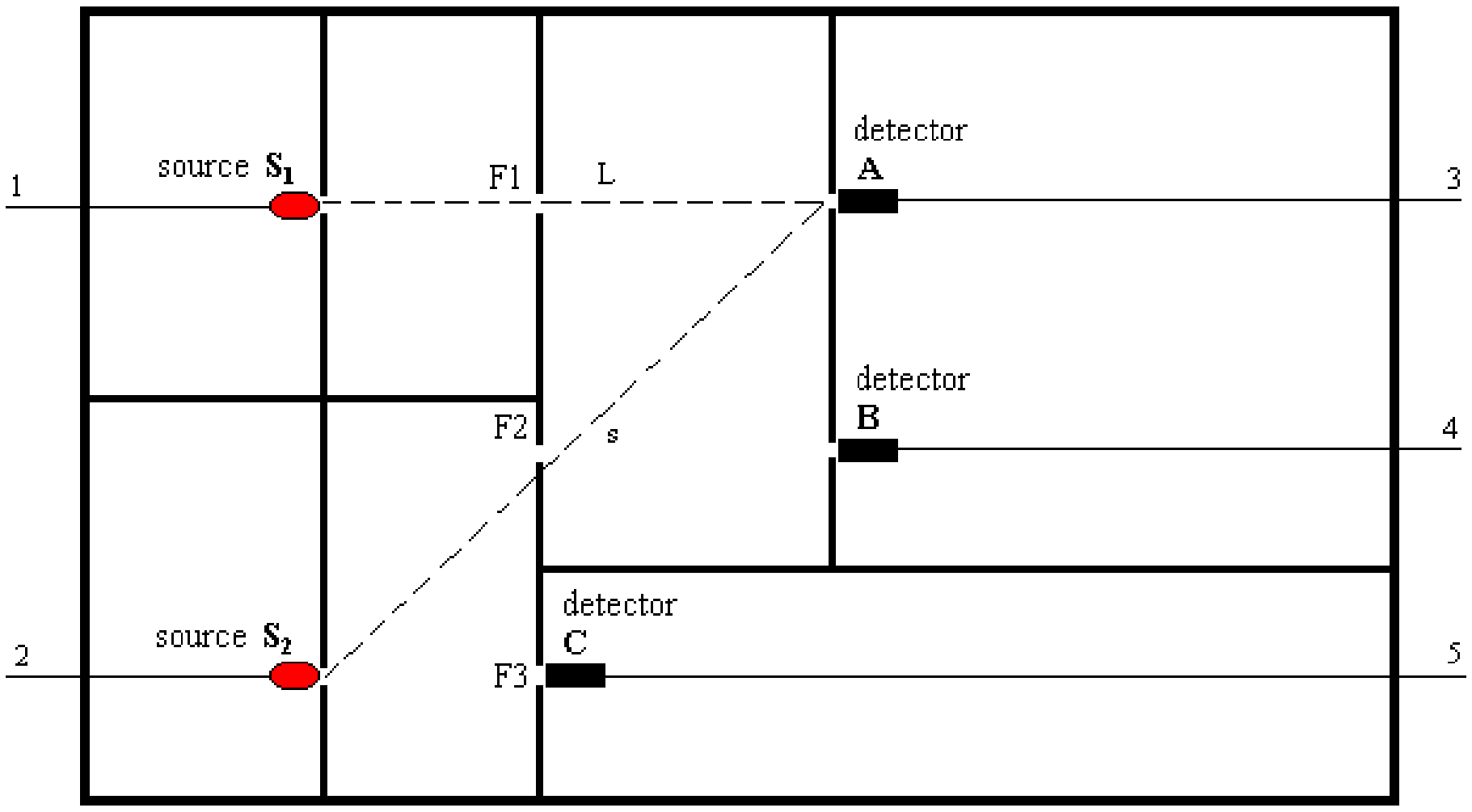

Fig. 1

The box (thoroughly screened from those frequencies which might have affected the measurements) contained two identical infrared (IR) LEDs, as (incoherent) sources of light, and three identical photodiodes, as detectors (A, B, C). The two sources $S_1$, $S_2$ were placed in front of a screen with three circular apertures $F_1$, $F_2$, $F_3$ on it. The apertures $F_1$ and $F_3$ were lined up with the two LEDs A and C respectively, so that the IR beams propagated perpendicularly through each of them. The geometry of this equipment was designed so that no photon could pass through the aperture $F_2$ on the screen. Let us stress that the dimensions of the apparatus were inferred from the geometrical size of the Florence microwave experiment [5], namely the horizontal distance between the planes of the antennas.

The wavelength of the two photon sources was $\lambda = 8.5 \cdot 10^{-5}$ *cm.* The apertures were circular, with a diameter of 0.5 *cm*, much larger than $\lambda$. We worked therefore in absence of single-slit (Fresnel) diffraction. However, the Fraunhofer diffraction was still present, and its effects have been taken into account in the background measurement.

Detector C was fixed in front of the source $S_2$; detectors A and B were placed on a common vertical, movable panel (see Fig. 1). This latter feature allowed us to study the space dependence of the anomalous effect.

Let us highlight the role played by the three detectors. Detector C destroyed the eigenstates of the photons emitted by $S_2$. Detector B ensured that no photon passed through the aperture $F_2$. Finally, detector A measured the photon signal from the source $S_1$.

In essence, the experiment just consisted in the measurement of the signal of detector A (aligned with the source $S_1$) in the two following states: (1) only the source $S_1$ switched on; (2) both sources on. Due to the geometry of the apparatus, *no difference in signal on A between these two source states ought to be observed, according to either classical or quantum electrodynamics.* The possible energy difference $\Delta_A$ in the signal measured by A when source $S_2$ is off or on (and the signal in B is strictly null) was considered evidence for the searched anomalous effect.

The measurement procedure was as follows:

**Step 1:** Measurement of the signal from detector A with source $S_1$ turned on and source $S_2$ turned off;
**Step 2:** Measurement of the signal from detectors A, B and C with $S_1$ off and $S_2$ on;
**Step 3:** Measurement of the signal from A, B and C with both sources $S_1$ and $S_2$ on.

The outcome of this first experiment was positive. The envisaged effect was observed indeed [1]. Precisely, the measured signal difference on detector A, $\Delta_A$ , ranged from $2.2 \pm 0.4$ $\mu V$ to $2.3 \pm 0.5$ $\mu V$. Moreover, such an anomalous effect was observed within a distance of at most 4 *cm* from the sources [1].

### *2.1.2. Second experiment*

The purpose of the second experiment (still carried out at L'Aquila in 2004) was to corroborate the results of the previous one [2]. The experimental set-up was essentially the same, but with a right-to-left inversion along the bigger side of the box and with three detectors of a different type from that used in the first investigation. In this way, it was possible to study how the phenomenon changes under a spatial parity inversion[2] and for a different type of detectors.

We used therefore the same Plexiglas box with wooden base and lid, containing two similar photon sources, three circular apertures and three similar phototransistors. The layout of the experimental set-up, seen from above, is shown in Fig.2 and is, of course, the mirror image of Fig. 1.

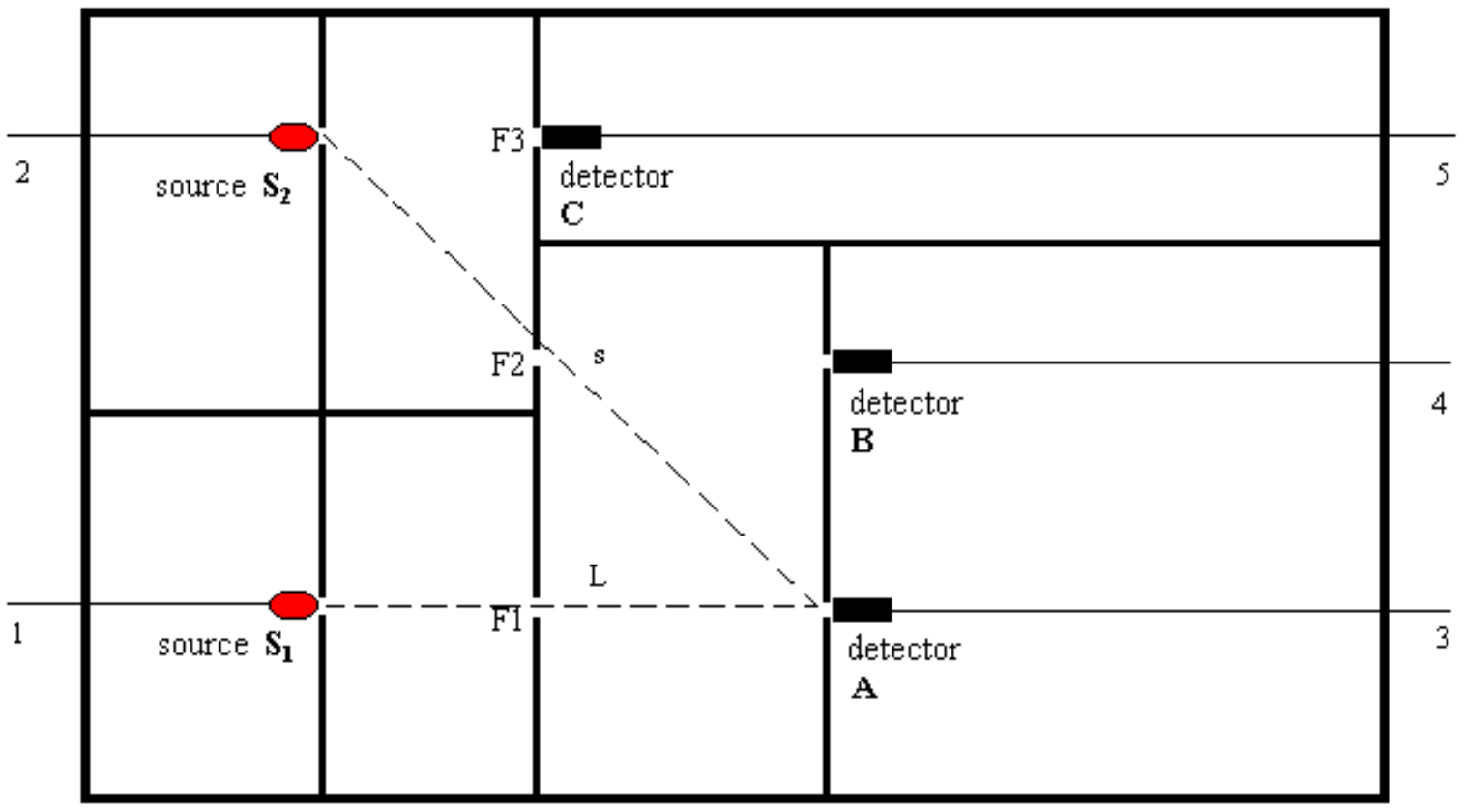

Fig. 2

[2] Testing the phenomenon under parity was suggested to us by Tullio Regge (private communication).

The dimensions of the apparatus, and the relevant quantities (like photon wavelength and aperture diameter), were identical to those of the first experiment.

The main difference with respect to the equipment of the first experiment was in the three identical detectors, which were not photodiodes but phototransistors (of the type with a convergent lens).

We expected to measure a signal difference on detector A, due to the breakdown of LLI, unlike the signal difference measured in the first experiment.
This is due to the fact that the detectors in the second experiment (phototransistors) had a lower "relative efficiency" compared with the detectors (photodiodes) used in the first one. One can define the relative geometrical efficiency ($\eta_g$) of the phototransistor (with respect to the photodiode) as the ratio of their respective sensitive areas, and their relative time efficiency ($\eta_t$) as the ratio of their respective detection times. Then, one can define the *relative total efficiency* ($\eta_T$) of the phototransistor with respect to the photodiode as the product $\eta_T=\eta_g\eta_t$. From the values of $\eta_g$ and $\eta_t$ in our case, one gets $\eta_T$=0.0015 [2].

Therefore, it was reasonable to presume that the value of the expected phenomenon to be given by the product of the total relative efficiency times the value measured in the first experiment, i.e. $\eta_T\cdot(2.3 \pm 0.5\mu V) = 0.004 \pm 0.001\ \mu V$. This was the difference in signal expected to be measured by detector A between the two source states, $S_1$ on, $S_2$ off (state 1); both sources on (state 2). This predicted value is two orders of magnitude below that measured in the first experiment. Therefore, we foresaw that the phenomenon would have been observable only at distance of 1 *cm* (corresponding to the position number 1 in the previous experiment, when we investigated also the spatial extension of the effect: see ref. [1]).
The measurement procedure followed the same three steps as in the first experiment (see Subsubsect. 2.1.1). Thirty groups of measurements were gathered (gaussian samples), and, by applying an inferential statistical test, we obtained the gaussian distribution of the minimum and maximum values for every measurement and each detector. The expectations of these two distributions (minimum values, maximum values) were computed for each detector. Then, we calculated the mean of these two expectation values (total mean $\overline{V}$). Let us denote by $\Delta\overline{V}$ the difference of the total averages, corresponding to the measurements on each detector for the two states of the sources. The experimental results are reported in Table 1 [2].

**Table 1 - Results of the second experiment.**

| | ***TOTAL MEAN*** $\overline{V}$ ***($\mu V$)*** | | | | | | ***DIFFERENCES BETWEEN TOTAL MEANS*** | |
|---|---|---|---|---|---|---|---|---|
| **Sources** | $S_1$ | $S_2$ | $S_1$ | $S_2$ | $S_1$ | $S_2$ | | |
| **State** | ON | ON | ON | OFF | OFF | ON | ***SIGN*** | ***ABSOLUTE VALUE*** $\Delta\overline{V}$ ***($\mu V$)*** |
| **Detector A** | 8.634±0.003 | | 8.626±0.003 | | | | (ON ON) > (ON OFF) | 0.008±0.003 |
| **Detector B** | 0.020±0.003 | | | | 0.275±0.003 | | (ON ON) < (OFF ON) | 0.255±0.003 |
| **Detector C** | 16.226±0.003 | | | | 16.476±0.003 | | (ON ON) < (OFF ON) | 0.250±0.003 |

The differences between the total means of the measurement on each detector for the two possible states of the sources allow us to draw the following conclusions.
The detector B was always underneath the maximum dark threshold that corresponds to 0.7 $\mu V$. The value of the difference for B had the same sign and the same order of magnitude of that of the detector C, which, conversely, was always exposed to radiation. Hence, we could speak of a common difference for the detectors B and C, which can be regarded as a *device* signal effect. The difference for detector A had an opposite sign with respect to that of detectors B and C and was lower by two orders of magnitude.
Since the detector B was always below the maximum dark threshold, it can be inferred that no photons from $S_2$ went through the aperture $F_2$[3]. Therefore, the disparity between the difference on detector A and those on detectors B and C *cannot be attributed to photons that passed through the aperture $F_2$*. As a consequence, we regard the difference on A *as a true signal difference* and not as a device effect.

Moreover, the value of the difference measured on detector A (0.008 ± 0.003 $\mu V$), is consistent, within the error, with the difference $\Delta_A \cong 2.3\ \mu V$ measured in the first experiment, *provided that the unlike efficiency between photodiodes and phototransistors is taken into account* (see the definition of the total relative efficiency between phototransistors and photodiodes, $\eta_T$, given above).

It is also worth noticing the difference of about one order of magnitude between the signals at detector B in the two source states (see next Section for a possible explanation of this fact).

The outcomes of this second experiment confirmed those of the first one for the following reasons:

i) Consistency of the measured value with that obtained in the first experiment, despite the new type of detectors employed;
ii) The sign of the signal variation on the detector A (that measures the phenomenon), which is always opposite to the signal variations on the two controlling detectors B, C;
iii) The difference of about two orders of magnitude between the variation of the signal on A and the variation common to B and C.

Moreover, the observed effect is apparently not affected by the parity of the equipment.

## 2.2. *Crossing photon beam experiments*

The results of our experiments suggest that similar anomalous effects can be observed also in different experimental situations involving photon systems, like e,g, in interference experiments. Further evidence for the anomalous photon system behaviour and for the anomalous photon-photon cross section was observed indeed in orthogonal crossing photon beams.

These interference experiments were carried out after our first one, one with microwaves emitted by horn antennas (see Fig.3), at IFAC – CNR (Ranfagni and coworkers) [7,8], and the other with infrared $CO_2$ laser beams (Fig.4), at INOA (Meucci and coworkers) [8]. Let us summarize the results obtained.

[3] In order to support this statement, we point out that the total mean of the signal on detector B when both sources are on is lower of one order of magnitude than that when only $S_2$ is on. Since, of the two sources, $S_1$ can affect more the response of B and switching it on makes the total mean of B decrease, we can infer that $S_2$ affects it much less. Thence, no photons from $S_2$ can go through $F_2$.

The main result of the IFAC experiment consists in an unexpected transfer of modulation from one beam to the other, which cannot be accounted for by a simple interference effect. This confirms the presence of an anomalous behaviour in photon systems, in the microwave range too.

In the optical experiment carried out at INOA-CNR [1,2,8], the wavelength of the used infrared laser beams was 10600 *nm*, namely one order of magnitude higher than the wavelength of the sources (LEDs) used in our experiments (850 *nm*). Let us also remark that the energy of the photons of our two experiments was $10^4$ times higher than that of the photons in the Cologne and Florence experiments [4,5,7], and 10 times higher than that of the INOA-CNR experiment [8].

The optimum alignment which can be achieved with lasers and the laser beam confinement make this optical set-up especially suitable for investigating the anomalous behaviour of the photon-photon cross section. The measurements were carried out for a relatively long lapse of time, that is, 12 minutes. This allowed one to perform a statistical test on the averaged results [2,8]. The signal statistics provided a significant variation in the mean values obtained with or without beam crossing. Hence the chance to have two identical statistics was rejected with a sufficient level of confidence. We evaluated [2] that the actual shift of the crossed beam signal with respect to the single beam signal is $2.08 \pm 0.13\ \mu V$. This value agrees excellently with that obtained in our first experiment $\Delta_A \cong 2.3\ \mu V$. Notice that *the laser experiment shows that the observed phenomenon does depend neither on the infrared wavelength, nor on the coherence properties of the light.*

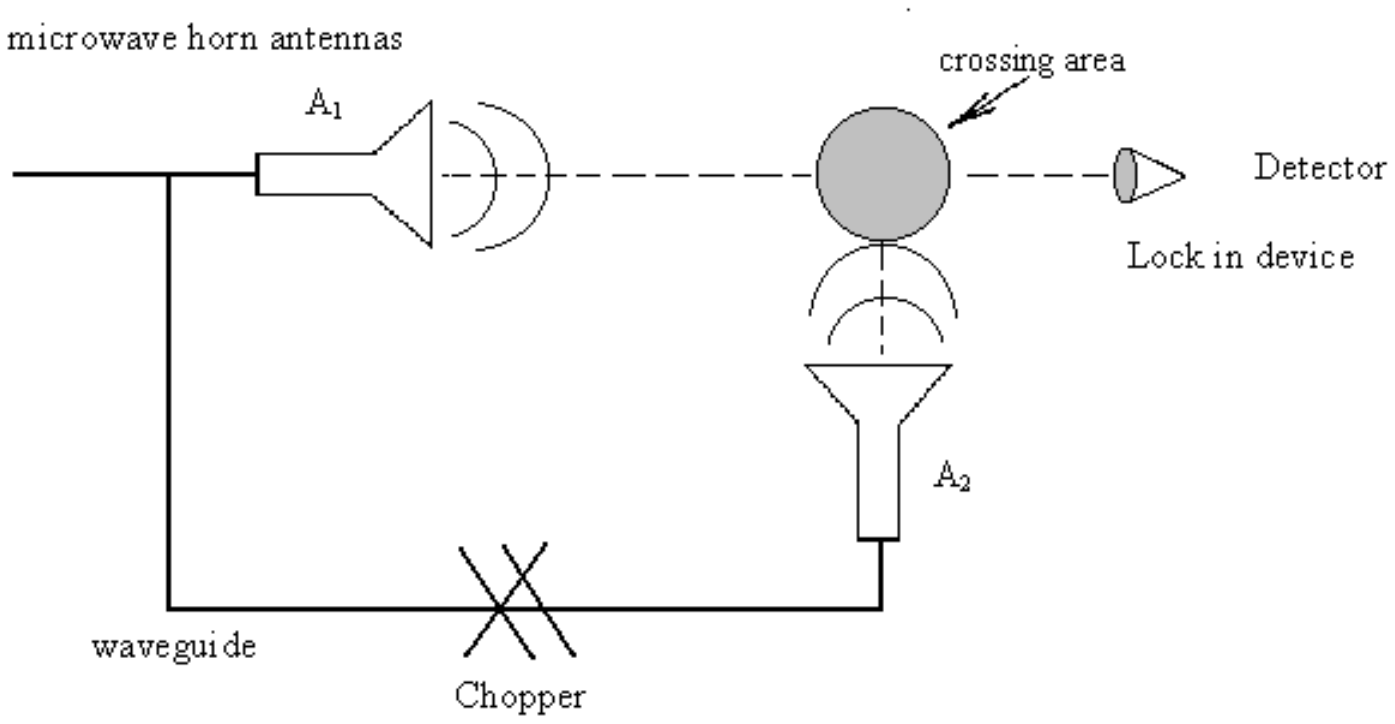

Fig. 3

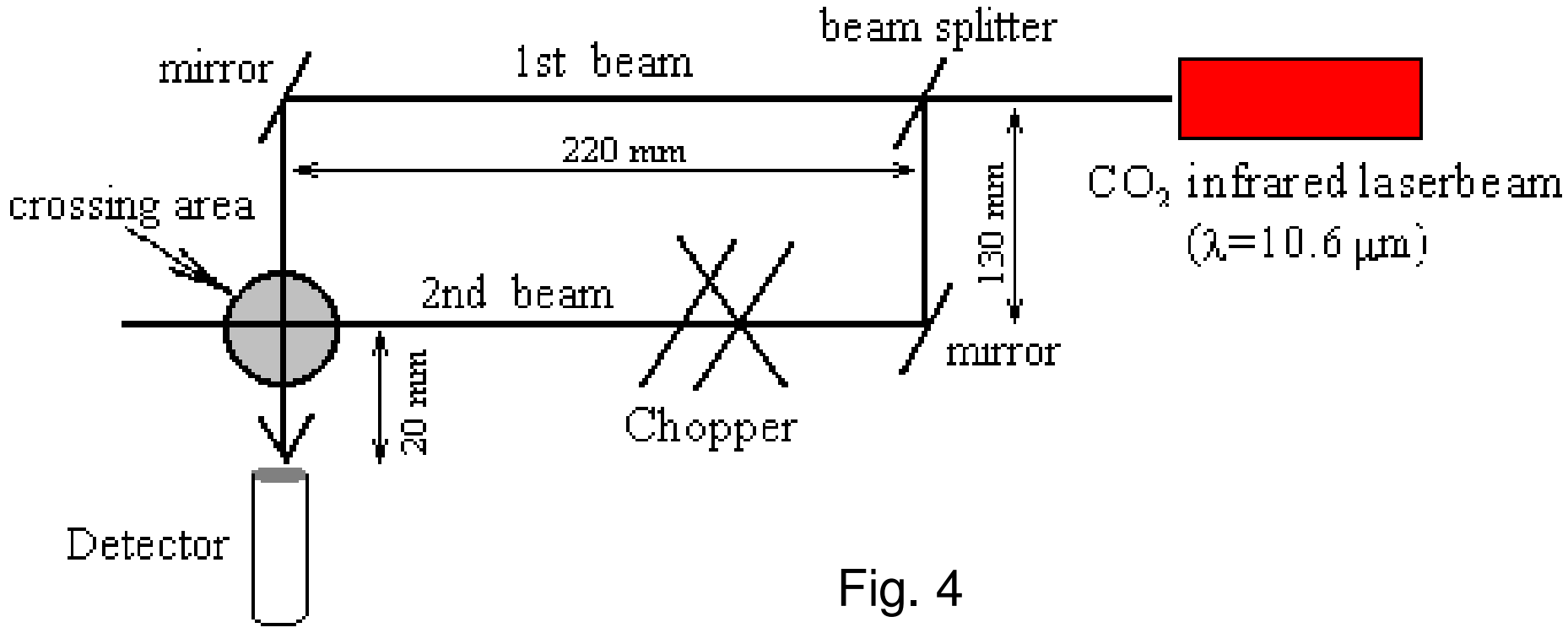

Fig. 4

# 3. Discussion and conclusions

### *3.1. The shadow of light: Hollow wave and LLI breakdown*

The observed anomalous behaviour of photon systems cannot be explained in the framework of the Copenhagen interpretation of quantum wave [1,2].

First of all, let us consider the difference $\Delta_A$ in the signal measured by detector A according to whether only $S_1$ is turned on or both sources are on, and recall the role played by the three detectors in both our experiments (see Subsect. 2.1). On one side, detector C measures – and hence destroys – the superposition of states belonging to the photons emitted by $S_2$ (thus manifesting their corpuscle nature); on the other hand, detector B is always underneath the dark voltage threshold, thus ensuring no transit of photons through aperture $F_2$ . Therefore in no way – according to the Copenhagen interpretation – photons from $S_2$ can interact with those from $S_1$ , thus accounting for the signal difference on detector A.

On the contrary, such a result can be understood by interpreting – following Einstein, de Broglie and Bohm – the quantum wave as a hollow wave.

In such a framework, pilot waves can interact with quantum objects (as assumed by de Broglie and Andrade y Silva [9]). Then, the region outside aperture $F_2$ is optically forbidden to the photons emitted by the source $S_2$ , *but not to the hollow waves associated to them*. Thence, the photons emitted by the source $S_1$ can interact with the hollow waves of photons from the source $S_2$, which have gone through the aperture $F_2$. Consequently, the change $\Delta_A$ in the A signal – in absence of any change in the response signal of detectors B and C – finds a natural explanation, in the Einstein-de Broglie-Bohm interpretation of quantum wave, in terms of the interaction of the $S_1$ photons (and their hollow waves) with the hollow waves (of $S_2$ photons) passed through $F_2$.

The role played by the aperture $F_2$ is fundamental, since, although hollow waves can penetrate in optically forbidden regions, nonetheless the mass distribution and density are expected to affect their propagation. Hence, they can pass only through space regions with a lower mass density.

Let us show that our results can be regarded as evidence for the breakdown of local Lorentz invariance too. The connection with the LLI breakdown is provided by the marked threshold behaviour the phenomenon exhibited. In fact, the anomalous effect was observed within a distance of at most 4 *cm* from the sources (1 *cm* in the second experiment), and the measured signal difference on detector A ranged from $\Delta_A \cong 2.3\ \mu V$ (first experiment) to $\Delta_A \cong 0.008\ \mu V$ (second experiment)[4] [1,2]. These values are consistent with the threshold behaviour for the electromagnetic breakdown of LLI, obtained in the framework of Deformed Special Relativity [3]. This is why we called *non-Lorentzian* the anomalous observed effect [1].

Since, according to DSR, the breakdown of LLI is connected to a deformation of the Minkowski metrics, this latter result supports the hypothesis (we first put forward in [1]) that *the hollow wave (at least for photons) is nothing but a deformation of space-time geometry, intimately bound to the quantum entity ("shadow of light").*

This can be depicted as follows. Most of the energy of the photon is concentrated in a tiny extent; the remaining part is employed to deform the space-time surrounding it and, hence, it is stored in this deformation. It is just the deformations ("shadows") of the photons from

[4] We recall that this second value of $\Delta_A$ is consistent, within the error, with that measured in the first experiment, if the total relative efficiency between phototransistors and photodiodes, $\eta_T$, is taken into account (see Subsubsect.2.2).

$S_2$ that expand, go through $F_2$ and interact with the shadows of the photons emitted from $S_1$.

Therefore, in this view, the difference of signal measured by the detector A in both our experiments can be interpreted *as the energy absorbed by the space-time deformation itself*, which cannot be detected by the central detector B[5]. In other words, our experimental device, used in both experiments, "weighed" the energy corresponding to the space-time deformation by the measured difference on the first detector.

If the interpretation we have given here is correct, *our experiments, among the others, do provide for the first time direct evidence for the Einstein-de Broglie-Bohm waves and yield a measurement of the energy associated to them.*

The hypothesis of the hollow wave as space-time deformation is able to explain also the anomalous behaviour observed in crossed photon beam experiments [7,8]. In fact, the shadow of the photon spreads beyond the border of the space and time sizes corresponding to the photon wavelength and period, respectively. This changes the photon-photon cross section (strongly depressed both in classical and in quantum electrodynamics)[6], and gives rise to anomalous effects in photon-photon interactions, like in crossing photon beams.

### *3.2. Action-at-a-distance and failure of relativistic correlation*

As already noted, our experimental equipment, in both experiments, was just sized according to the results (analysed by means of the DSR formalism [6]) of the Cologne [4] and Florence [5] experiments, which evidenced superluminal propagation of electromagnetic signals. This implies that, in all these kinds of experiments, *the relativistic correlation* (in the Einstein's sense, i.e. as correlation at the light speed c) *fails*. This agrees with the geometrical interpretation within the DSR framework, according to which the photon propagation occurs in a deformed space-time, where *c* is no longer the maximal causal speed [3].

We can conclude that it is just the de Broglie-Bohm hollow wave, seen as "the shadow of light" – namely as a space-time deformation, breaking LLI, which affects quantum objects in seemingly inaccessible, far regions –, the true responsible of the failure of relativistic correlation. In this sense, it apparently represents an action-at-a-distance without any transport of energy (which Einstein, within the domain of Quantum Mechanics, called "spooky action at a distance"). However, just our interpretation of the hollow wave as a space-time deformation which moves together with the quantum object – the photon in this case – is actually an indication of the opposite view. As a matter of fact – as discussed above –, part of the photon energy is detected by a direct measurement of photons by the third detector C; the remaining part is used to deform the space-time of every photon and it is evidenced by the difference measured by the first detector. Hence, it is no longer correct to say that there exists (at least in this framework) an action-at-a-distance without any energy transport.

[5] One might think to detect such an "energy of deformation field" (corresponding to the hollow waves of photons) by a detector operating by the gravitational interaction, rather than the electromagnetic one. However, this would still be impossible, because the deformation value lies within the energy interval for a flat (Minkowski) gravitational space-time, according to DSR [3]. We are deeply indebted to G. Caricato for this and other precious remarks on the topics of action-at-a-distance in our experiments (correspondence between Caricato and F. Pistella).

[6] In fact it goes as $\alpha^4$ (with $\alpha$ being the fine structure constant).

### 3.2. *Failure of Bohr's principle of complementarity*

A careful analysis seemingly shows that the outcomes of our experiments, in particular those of the second one, invalidate the Bohr principle of complementarity too. Let us explain why.

As is well known, the probabilistic meaning of quantum wave, in the Copenhagen interpretation of the formalism of quantum mechanics, implies that a quantum entity, like a photon, is a superposition of eigenstates. Therefore, in this sense, it is neither a wave nor a corpuscle. Only a measurement can give evidence to the wave nature or to the particle nature of the quantum object, and these features appear under mutually exclusive experimental arrangements. This last sentence is the essential content of Bohr's complementarity principle.

Then, in this view, particle and wave are two complementary properties of a quantum entity. Position is a "particle-like" property, and when a photon, or an electron, makes a spot of light on a screen, we know exactly where it is but we do not know how it got there. Conversely, a wave is a spread-out thing with no well-defined position, but with a well-defined direction of motion.

We will show that this is not the case in our experiments, by analysing their results (in particular those of the second one, reported in Table 1), on the basis of the geometry of the apparatus and of the emitting and detecting features of the sources and the detectors, respectively.

Let us stress once again that the apparatus of both our experiments was designed according to geometrical optics. We recall that the employed sources had an angular aperture of the emitted power of 20°, and that the response of the detectors was null for angles wider than about 10° [1,2]. Both the distances between sources and apertures, and those between adjacent apertures, were designed so that the photons, emitted by $S_1$, could only pass through aperture $F_1$ and be detected only by detector A. Analogously, photons, emitted by $S_2$, could only go through aperture $F_3$ and hence be detected only by detector C.

Let us concentrate on the latter, i.e. on photons emitted by $S_2$. According to the above discussion, we know exactly where these photons began their flight (source $S_2$) and exactly where they ended up to (detector C). In other words, our experimental arrangement was designed in order to give evidence of the corpuscle nature of the photons emitted by $S_2$ (and analogously for $S_1$). There was an initial position, a final position and a trajectory too, since, had we placed other $F_3$-like apertures between $S_2$ and C, the response of detector C would not have changed.

On the other hand, detector B ensures that no photons from $S_2$ passed through $F_2$ (ruling out the ambiguities in the framework of path integrals). We can therefore conclude that, in principle, *our apparatus would have to work exactly according to the laws of geometrical optics, with photons following well-defined, geometrical trajectories (and therefore exploiting their corpuscular nature).*

However, a glance at Table 1 with the results of the second experiment shows that actually the behaviour of the system is more complex, and in some respects contradictory. As already stressed, the mere fact that $\Delta_A \neq 0$ invalidates the probabilistic interpretation of the quantum wave. However, also the response signal of detector B exhibits an anomalous behaviour. In fact, when both sources are on (state 2), and therefore more photons are present in the system, the signal on A (total mean) increases with respect to state 1, when only the source $S_2$ is on (and therefore a lower number of photons is present in the system). This behaviour of A is in principle a correct one, although unaccountable in terms of probabilistic wave (as already stressed). The signal on C remains constant (within the device working fluctuations). On the contrary, *the total mean of the signal on B in state in state 2* ($S_1$ on, $S_2$ off) *is lower by an order of magnitude than that in state 1* ($S_1$ off, $S_2$ on) (see Table 1). We want to stress here again that this experimental evidence gives strength to what is already ensured by both of the total means being below the dark threshold, that is no photons go through F2. This strengthening role, played by this evidence, becomes clear once it is noted that, because of the disposition of the sources and the apertures, photons from $S_1$ can affect the response of detector B more than those from $S_2$[7]. In spite of this, when $S_1$ is turned on ($S_2$ being already on) the total mean of the signal on B decreases.

This anomalous behaviour of detector A and detector B can be still explained on the basis of the hypothesis of the shadow of light, by interpreting it as manifestations of *an anomalous interference phenomenon involving hollow waves.*

In this view, the anomalous signals at detectors A and B are regarded as part of an interference pattern. Precisely, we interpret the *higher* signal on A when both sources are on as a *constructive interference* of the $S_1$ photons with the space-time deformations (hollow waves) associated to the $S_2$ photons, which yields a bright fringe. Analogously, the *lower* signal on B (that always "sees" the dark) in the same system state 2 (when both sources are on) is considered as a *destructive interference* of the space-time deformations belonging to the photons from $S_1$ and $S_2$ with those belonging to the photons of the dark. Because of the slenderness of the interference fringes (remember that in the second experiment is $\Delta_A \cong 0.00.8\ \mu V$), we emphasize that these phenomena can be discerned only as ensemble or multi-quantum processes.

Needless to say, the anomaly of such an interference behaviour lies in its occurring in the framework of purely geometrical optics (as stressed above), in which incoherent sources cannot interfere. Moreover, in our opinion it is strictly related to the breakdown of local Lorentz invariance, consequent to the threshold behaviour of the effect (in particular, to being $\Delta_A \leq E_{0,e.m.}$ , with $E_{0,e.m.}$ the energy threshold for LLI breakdown derived in the DSR framework [3]). Let us recall that it is just this behaviour that allowed us to regard the hollow wave as a space-time deformation. This is why we shall refer to such an anomalous interference, involving space-time deformations associated to quantum objects, as *non-Lorentzian interference.*

Then, we can conclude that, in our experiments*, a wave behaviour is present within a purely optical-geometric setting.* Indeed, as already stressed, our equipment was strictly designed to work according to the laws of geometrical optics. The corpuscular aspect is evidenced by the fact that the responses of the three detectors A, B, C ensure the right geometrical connection between $S_1$ and A, on one side, and between $S_2$ and C, on the other side. However, we have just seen that the different signals measured by A and B in the two different states of source switching can be interpreted as evidence of interference among the hollow wave of photons, regarded as space-time deformations. Therefore, *both corpuscle and wave properties manifest themselves in the same photon system, at the same time, in the same experimental apparatus.* Consequently, *our experimental results seemingly invalidate Bohr's principle of complementarity.*

Let us notice that recently Afshar carried out a double-slit experiment, that apparently proves a failure of the complementarity principle [10]. He essentially showed that the coherent superposition state, corresponding to the interference pattern, persists regardless of the fact that the which-way information (trajectory) is obtained in the same experimental apparatus. Moreover, he states that evidence for coherent wavelike behaviour is not a single-particle property, but an ensemble (multi-particle) property.
Although the results of ref. [10] do agree with ours, in denying the validity of the wave-courpuscle complementarity, let us stress that Afshar's experiment is in a sense *dual* (or complementary) to ours. In fact, in [10] a corpuscle behaviour (trajectory) was observed within a pure wave setting (double-slit interferometer); on the contrary, in our case a wave behaviour (interference) was observed within a pure corpuscular framework[8].

[7]In fact, the intensity of photons from $S_1$ , going through $F_1$ - which might be diffracted by the aperture edge towards B - is much greater than that of photons from $S_2$ , going through $F_2$. Actually, this latter intensity must be zero, according to the purely optical-geometric setting of the equipment.

***Acknowledgements***

It is both a pleasure and a duty to warmly thank Gaetano Caricato and Tullio Regge, to whom we are strongly indebted for their deep review and highly useful remarks. We are also grateful to Riccardo Meucci for communicating to us the preliminary results of his experiment. Stimulating discussions with (and useful criticism by) Basil Hiley, Eliano Pessa and Anedio Ranfagni are gratefully acknowledged. Last but not least, a special thank is due to the President of CNR, Fabio Pistella, for his kind interest in our work and continuous encouragement.

[8] Notice that the grid in Afshar's experiment plays a role analogous, but dual, to that of detector B in ours. In fact, the grid assures the wave nature of the photon without affecting its particle features, whereas the detector B works as a veto for the photon as a particle without affecting its wave properties.

## FIGURE CAPTIONS

**Fig. 1** Above view of the experimental apparatus used in the first experiment.

**Fig. 2** Above view of the experimental apparatus used in the second experiment; note that it is the mirrored image of Fig. 1.

**Fig. 3** – Crossed-beam experiment in the microwave range, exploiting two horn antennas.

**Fig. 4 –** Crossed-beam experiment in the infrared range, exploiting a $CO_2$ laser.